\documentclass{IEEEcsmag}

\usepackage[colorlinks,urlcolor=blue,linkcolor=blue,citecolor=blue]{hyperref}
\usepackage{upmath}
\usepackage{multirow}
\usepackage[table,xcdraw]{xcolor}
\usepackage{blindtext}
\usepackage[justification=centering]{caption}
\usepackage{comment}
\usepackage{booktabs,multirow}
\usepackage{multicol}

\jvol{XX}
\jnum{XX}
\paper{8}
\jmonth{May/June}
\jname{IT Professional}
\pubyear{2019}

\begin{document}


\title{Evaluation of Security Training and Awareness Programs: Review of Current Practices and Guidelines}

\author{Asangi Jayatilaka}
\affil{CREST – the Centre for Research on Engineering Software Technologies, School of Computer Science,
The University of Adelaide, Australia}

\author{Nathan Beu}
\affil{School of  Psychology, University of Adelaide, Australia}

\author{Irina Baetu}
\affil{School of Psychology, University of Adelaide, Australia}

\author{Mansooreh Zahedi}
\affil{School of Computing and Information Systems, The University of Melbourne, Australia}

\author{M. Ali Babar}
\affil{CREST – the Centre for Research on Engineering Software Technologies,
School of Computer Science,
The University of Adelaide, Australia}

\author{{L}aura Hartley}
\affil{Independent Consultant, Melbourne, Australia}

\author{Winston Lewinsmith}
\affil{Independent Consultant, Melbourne, Australia}

\markboth{Department Head}{Paper title}


\begin{abstract}
Evaluating the effectiveness of security awareness and training programs is critical for minimizing organizations' human security risk. Based on a literature review and industry interviews, we discuss current practices and devise guidelines for measuring the effectiveness of security training and awareness initiatives used by organizations.  
\end{abstract}

\maketitle


\chapterinitial{T}he exponential growth of digitalization has led to significant growth in security threats and risks. No matter how small they are, businesses now need to be aware of and consciously take appropriate security measures at every level.  Despite the availability of many countermeasures, technologies, and solutions to thwart security attacks, successful security attacks are on the rise. Technology alone cannot be considered a comprehensive solution to the increasingly sophisticated security threats; the people in an organization are actually the primary target and the weakest line of defense~\cite{bauer2017prevention}. When it comes to the “human factor”, security training and awareness programs are critical for maintaining up-to-date knowledge of security risk-prone activities and available mitigation measures. The goal of these programs is to  change employee behaviors positively by educating staff about organizational security policies and safe security practices. An engaging security awareness training program helps organizations covert their employees into their first line of defense.

Security training and awareness in and of itself is not enough; it is essential to measure the effectiveness of those initiatives  \cite{bada2019cyber}. Those who are tasked with managing awareness programs often question, “how can we show that security awareness training is worth the investment?” and  “how can we show that security awareness training is creating behavioral changes in employees?”. Adequate evaluation of these training programs allows organizations to gauge what aspects of the training improve performance and what aspects could be improved. Unfortunately, it is still unclear which metrics yield the most information and how to  measure the effectiveness of the security training and awareness programs comprehensively \cite{bada2019cyber}.

Metrics are quantitative assessments of processes or performance, and the choice of relevant metrics plays a crucial role in any measurement program. The importance of careful choice of the relevant metrics is highlighted in the ``Security Awareness Maturity Model” developed by the SANS institute \cite{SANS}. The model has 5 phases with ``non-existent" being the lowest level and ``metric framework" being the highest level. In a  global survey conducted with over 1,500 security professionals across 91 countries, the majority of the respondents (53\%) reported that their programs fall in the middle in terms of the SANS maturity levels (i.e., promoting awareness and behavior change) and only a minority ($<$7\%) reported that they have a metric framework in place for measuring their security awareness program \cite{SANS}. The survey results re-emphasize the gap between security awareness and training delivery and measurement. Without adequate measurement of a program's effectiveness, one cannot know how well it works, whether it's worth the cost, or what to improve. The goal of this work article is to address this gap by reviewing the existing practices and providing guidelines for comprehensive security awareness and training program evaluation.

To this end, we reviewed the existing literature related to: i)~metrics used for measuring the effectiveness of security training and awareness programs; and ii)~behavioral and cognitive principles to design, implement, and evaluate the efficacy of security training and awareness programs. Furthermore, we carried out a series of interviews with the Security Awareness team of a leading Australian financial institute to obtain in-depth insights into the existing practices, tools and metrics they use for measuring the effectiveness of security awareness and training programs. These interviews allowed us to discuss the security training and awareness programs they carry out and also to analyze the tools and metrics they use to evaluate their programs.
Based on the gathered knowledge we:

\begin{itemize}
    \item Discuss the existing practices and metrics for gaining insights into the effectiveness of security training and awareness programs with an emphasis on practices and metrics for analyzing employees' behavioral changes.
    \item   
    Discuss drawbacks and/or considerations related to several identified  practices and  metrics and provide suggestions on how organizations can gain a better understanding of the behavioral changes resulting from security training and awareness programs.  
    \item Provide guidelines for comprehensive evaluation of security training and awareness programs carried out by organizations.

\end{itemize}

\section{Practices and metrics for measuring effectiveness of security awareness and training programs}

The existing key practices and metrics we identified for measuring the effectiveness of training and awareness programs can be  mapped into the three levels: i)~reactions; ii)~learning; and iii)~behaviors in the Kirkpatrick Model \cite{kirkpatrick2006evaluating} .  

The Kirkpatrick Model's first level measures the learners' reaction to the training. It is understood that a strong correlation exists between learning retention and how much learners enjoyed their learning experiences and whether they found them valuable. Employees' reactions to training are usually obtained immediately on the completion of training using surveys. The second way to understand the effectiveness of security training and awareness programs is through measuring employee security knowledge and awareness. Multiple or periodic measurements of users' security awareness (e.g., using surveys \cite{parsons2017human}) may provide insights into how effective the training and awareness programs are in terms of providing security awareness and knowledge to employees \cite{parsons2017human}.   

Unfortunately, awareness or knowledge alone will not change an organization's risk profile. Behavioral scientists have demonstrated that in most cases, people make decisions based on intuition, emotion, and social pressure, and not based  on knowledge  alone \cite{bada2019cyber}. Therefore, the third level in the Kirkpatrick Model is focused on measuring employee's behavioral changes. Measuring employee behavioral changes can be tricky; however, if measured correctly, it can provide a deeper understanding of the effectiveness of an organization's security training and awareness program. 

Now we report on the state-of-the-art practices and metrics for measuring behavioral changes of employees. These are  grouped under five focus areas.  We also provide the details of a set of additional measures and information that can provide insights to better interpret the identified key metrics (see Tables 1, 3, and 4-6 for more details). Furthermore, we discuss the limitations and considerations with respect to the existing practices and provide suggestions as to where to improve (in Tables 1, 3, and 4-6,  these suggestions  are marked as `No' under the `Existing' column). Where a metric can be reported under more than one focus area, we have selected the best-matched focus area for presenting that metric.

\begin{table*}[!h]
  \caption{Phishing simulations: metrics and measures to identify behavioral changes of employees}
  \label{keymetrics-phishing}
  \label{tab:observationdetails}
  \centering 
  \resizebox{1\textwidth}{!}{%
  \begin{tabular} { p{.025\textwidth} p{.25\textwidth} p{.55\textwidth}  p{.045\textwidth}   p{.03\textwidth} }
    \toprule
   ID & Name & Description & Existing? & Risk$^*$  \\
    \midrule
    \multicolumn{4}{c}{\textbf{Key metrics}} \\
    \midrule 
   P1 & Click rate & The percentage of employees who clicked on phishing email links in  base campaigns & Yes &   \textcolor{red}{P} \\
   P2&   Re-click rate & The percentage of employees who fall for  phishing emails in the clicker campaigns. The clicker campaign considers only the employees who click on links in the base campaign.   & Yes &  \textcolor{red}{P}\\
   P3 &  Report rate  & The percentage of employees who reported the phishing emails & Yes & \textcolor{green}{N} \\
   P4 & Non-responder rate  & The percentage of employees who did respond to the phishing emails in anyway & Yes &  \textcolor{red}{P} \\
    
    P5 & Download rate & The percentage of employees who downloaded attachments in phishing emails & Yes  & \textcolor{red}{P} \\
    P6 &  Reply rate & The percentage of employees who replied to  phishing emails & Yes &  \textcolor{red}{P}  \\
   P7 & False positive rate${^*}{^*}$  & The percentage of employees who misjudged a legitimate email and reported that as a phishing email & Yes &   \textcolor{red}{P} \\
   P8 & Time taken to report  & The average time  taken to report a phishing email & Yes &  \textcolor{red}{P}  \\
    
\\
    P9 &   Likelihood of an employee to fall repeatedly  for phishing emails  &  Likelihood of an  employee of the organization to fall for phishing emails in the base campaign and in the subsequent re-phisher campaign. This is a multiplication of the click rate and re-click rate.
     & No &   \textcolor{red}{P} \\
      P10 &  Conversion of clickers to reporters &   Percentage of the clickers in the base campaign who converted to reporters in the subsequent phishing campaign
     & No &    \textcolor{green}{N} \\
     P11& Sensitivity Index ${^*}{^*}$ & Tells us how accurately an employee can identify a phishing email from a legitimate email, while controlling for threat sensitivity &   No &   \textcolor{green}{N}  \\
     P12 & Criterion Index ${^*}{^*}$  & Tells us an employee’s predisposition towards or away from the conclusion that an email is a phishing attempt, or sensitivity to threat. Informs us of their starting point for making decisions about the safety of an email, which can be informed by perceived organizational risk/reward for not detecting a phish. & No &   \textcolor{green}{N} \\
  P13 &    Confidence calibration &  The ability to correctly identify one's own level of performance and appropriately assign confidence to one's decision  & No &   \textcolor{green}{N} \\
    \midrule
    \multicolumn{4}{c}{\textbf{ Additional measures and information to better interpret the key metrics}} \\
    \midrule 
     P14 & Phish scale & A measure that recognizes the difficulty of a phishing email. & Yes   & N/A\\
    p15 & Viewer rate  & The percentage of employees who viewed the phishing email.  & Yes   & N/A \\
      P16 & Reasons for clicking and not clicking  & Follow-up with the clickers and non-clickers to understand their perceived reasons for clicking or not clicking.  & Yes & N/A\\
     P17 & Email response based on training completion &  A comparison between the email response (e.g., clicking, reporting, not responding) for those who completed training vs those who did not  & Yes & N/A\\
    P18 &  Completion of training assigned for re-clickers &   Percentage of clickers  who have completed  training  & Yes & N/A \\
     P19 &   Standard deviations across departments for phishing behaviors & Standard deviation of click rates, report rates and non-responder rates across departments & No & N/A \\
       P20 & Diffusion of responsibility & The extent to which an employee takes personal responsibility for the goals, outcomes, or duties of a group/organization compared with expecting them to be fulfilled by other members of the group/organization & No &   \textcolor{red}{P}\\
    \bottomrule
\end{tabular}}
\smallskip

\parbox[t]{\textwidth}{$^*$If the increase in the metric value  is associated with a higher  security risk  then we denote this as  \textcolor{red}{P}. If the increase in the metric value is associated with a lower  security risk  then we denote this as \textcolor{green}{N}. N/A denotes where such direct interpretation is not possible.}
\parbox[t]{\textwidth}{${^*}{^*}$Phishing simulation would need to include both phishing and legitimate attempts in order to be able to calculate this metric.}
\end{table*}

\begin{table*}[ht]
\centering
\captionsetup{justification=centering}
\caption{Four potential outcomes with respect to reporting a received  email}
\begin{tabular}{l|l|l|l }
   \toprule
\multicolumn{2}{l}{} & \multicolumn{2}{c}{\textbf{Phishing attempt}} \\ \cline{3-4} 
\multicolumn{2}{l}{\multirow{-2}{*}{}} & \textbf{Present} & \textbf{Absent} \\ \hline
\multicolumn{1}{c|}{} & \textbf{Report} & { \begin{tabular}[c]{@{}l@{}}Optimal desired response\\ (Correct identification)\end{tabular}} & { False positive} \\ \cline{2-4} 
\multicolumn{1}{c|}{\multirow{-2}{*}{\textbf{Action taken}}} & \textbf{Did not report} & { \begin{tabular}[c]{@{}l@{}}Sub optimal or undesired response \\ (False negative)\end{tabular}} & {\begin{tabular}[c]{@{}l@{}}No response \\ (Correct disregard)\end{tabular}} \\ \bottomrule
\end{tabular}
\label{phishing} 
\label{metrictable}
\end{table*}

\subsubsection{Phishing simulation}
Growing attention is being paid to intervention programs aimed at training users to identify and report phishing emails correctly. Organizations commonly use phishing simulation tools such as KnowBe4 and Cofense. Often phishing simulations are designed to train users in ``teachable moments" that occur when a mistake is made. For instance, if a participant clicks on a link in a phishing email sent from a simulation tool (during a phishing campaign), they are immediately presented with an intervention designed to train them not to fall prey to phishing attacks. Then, those employees are re-phished to determine whether or not they benefited from the learning from the intervention program. 

Phishing simulation programs collect information related to phishing campaigns that allow organizations to calculate various key metrics, such as the click rate, re-click rate and report rate, to understand their workforce's phishing-related behaviors (see Table~\ref{keymetrics-phishing}). Furthermore, there are other measures (e.g., the phish scale that can be used to quantify the difficulty of an email~\cite{steves2019phish}) reported in the literature that allows better interpretation of  the key metrics in Table~\ref{keymetrics-phishing}. 

Although the current state-of-art practices related to phishing simulations are useful in obtaining much-needed insights into phishing related behaviors of employees, they are limited in their capacity to provide in-depth understanding as to why those behaviors manifest, how they vary across departments, how they are changed by training, or whether behavior changes from base to re-clicker campaigns. We provide suggestions about a set of additional metrics and measures (e.g., P9 - P13  and  P19 - P20 in Table~\ref{keymetrics-phishing}) that organizations can utilize to obtain these in-depth insights. We describe some of these proposed metrics in detail below.

The existing phishing-related metrics do not provide in-depth insights into why an email was judged to be safe or unsafe or why it was or was not reported. These often-overlooked questions are crucial for interpreting an organization’s phishing-related metrics more effectively  and planning for any future training. To make inferences about such questions, we need to consider the four potential outcomes of receiving an email, represented in Table~\ref{phishing}. Stockhardt~\cite{stockhardt2016teaching} highlights the importance of considering the true behaviors associated with each cell in Table~\ref{phishing} for articulating and understanding the human behaviors involved in cybersecurity and naturalistic as well as embedded anti-phishing contexts, so that training can be targeted to at-risk individuals or groups. At-risk individuals and groups are not only those at risk of falling victim to phishing attacks but also those who are less likely to report phishing attempts and, therefore, represent a risk to organizations.

In instance where a correct identification (Present-Report) is inferred, an employee may over-report in all instances (including where a phishing attempt was absent; that is, false positives), and this report is, therefore, a reflection of poorly calibrated sensitivity (over-sensitive to threat). Over-estimation of threat is associated with  unnecessary cost to organizations and an eventual reduction in willingness to report over time.
In instances where a false positive (Absent-Report) is inferred, an employee may erroneously report or may exhibit heightened sensitivity or bias towards the expectation of phishing attempts. 
In instances where a false negative (Present-Does not report) is inferred, an employee may correctly identifying a phish, but may not report it. Some evidence suggests this behavior might be due to diffusion of responsibility, which, in this case, is the belief that others will report or have reported it, hence, there is no need to report it.  
The diffusion of responsibility can contribute to non-reporting behavior; the scores for this aspect can identify those employees who represent a higher risk due to a lower likelihood of reporting phishing emails. The data required to calculate diffusion of responsibility can be collected by administering an empirically validated questionnaire, potentially at training.

In instances where there is no attempt or report (Absent-Does not report), an employee may not be confident in his/her judgement, or indeed have perceived a threat but chosen not to report. Thus, tracking non-simulated emails or simulated emails without a pseudo-phishing attempt gives  the ability to calculate useful metrics that take into account all of these possibilities by computing a sensitivity index that reflects the ability to discriminate phishing from non-phishing emails. 
Signal detection theory \cite{stanislaw1999calculation} allows us to measure the ability to differentiate between the pieces of information that contain useful or informative patterns from those that do not. In this context, sensitivity can be conceptualized as an employee’s ability to distinguish between phishing and legitimate emails (i.e., signal and noise, respectively, in signal detection terms), where greater sensitivity indicates a greater ability to discriminate between legitimate and phishing emails. Bias refers to an employee’s tendency to treat an email as legitimate or phishing under uncertainty (or low confidence, discussed below), where a negative value reflects a bias toward the perception of phishing and a positive value reflects a bias toward the perception of legitimacy.

Unlike signal detection theory, confidence ratings may provide insights into reporting behavior. Confidence and confidence calibration are meta-cognitive principles that describe the degree of confidence, or certainty one has in one's judgements, and the accuracy of those judgements \cite{hosseini1982detectability}. Capturing this behavior in a simulated or experimental setting is often undertaken using such principles and a measure of employee confidence (e.g., “how confident were you in your judgement?”). In combination with an estimate of accuracy obtained using signal detection theory, confidence ratings, therefore, allow one to determine whether employees are aware of their own level of performance, that is, whether they realize the extent to which they are capable of detecting phishing attempts (this is called confidence calibration). Such a measure allows an organization to discriminate between employees who are accurate and confident (well-calibrated) and inaccurate and confident (i.e., overconfident), and importantly, accurate and not confident (i.e., under-confident), as identifying this cohort allows an institution to support behavior. In the latter case, employees may accurately detect a phishing attempt but  not report it due to a lack of confidence. In such instances, additional training to improve detection accuracy would not be beneficial; instead, training that aims to calibrate confidence and accuracy would achieve better results.

\begin{table*}[!h]
  \caption{Security champions: metrics and measures to identify behavioral changes of employees}
  \label{keymetrics-champions}
  \label{tab:observationdetails}
  \centering 
  \resizebox{1\textwidth}{!}{%
 \begin{tabular} { p{.025\textwidth} p{.25\textwidth} p{.55\textwidth}  p{.045\textwidth}   p{.03\textwidth} }
    \toprule
   ID & Name & Description & Existing? & Risk $^*$  \\
    \midrule
    \multicolumn{4}{c}{\textbf{Key Metrics}} \\
    \midrule 
     C1&    Employee consultations  &  The number of times employees consult  security champions & Yes &   \textcolor{green}{N}  \\
   C2 & Security behaviors with and without champions &  Comparison of the   security awareness  metrics (e.g., click rates) in departments or offices that do have champions vs. those that do not. &    Yes &  N/A \\

   C3 &  Champion density vs.  security behaviors & Correlation of coefficient can be used to measure the strength and direction of  linear relationships between champion density in departments and respective security  behaviors (e.g., report rate).   &   No & N/A \\
     
    \midrule
    \multicolumn{5}{c}{\textbf{Additional measures and information to better interpret the key metrics}} \\
    \midrule 
  
     C4 &  Reasons for employees to engage/not engage with security champions &  Reasons for employees  to engage/not to engage  with   security champions within their organizational units.  & No & N/A\\
      
    \bottomrule
\end{tabular}}
\smallskip

\parbox[t]{\textwidth}{$^*$If the increase in the metric value is associated with a higher security risk then we denote this as \textcolor{red}{P}. If the increase in the metric value is associated with a lower security risk then we denote this as \textcolor{green}{N}. N/A denotes where such direct interpretation is not possible. }
\end{table*}

\subsubsection{Security champion programs or ambassador programs}
Persuading an entire organization to adopt a “security first” mindset is not an easy task. This may be especially challenging for organizational teams that work in silos and are not well-equipped to communicate their messaging more broadly within the organization. Security champions more easily can reach employees across the entire organization and consistently communicate  security messages. 

Table~\ref{keymetrics-champions} illustrates various ways to measure employees' security behavior related to security champions. For example, higher levels of employee initiated consultations with security champions (C1) indicate employees' interest in organizational security and trust towards the security champions. 
Calculating such measures can be challenging if there are no protocols to obtain the relevant information from security champions and employees.
Furthermore, we point out the importance of looking beyond  the frequency of engagement with security champions alone and the necessity for exploring reasons   why  employees engage do  an do not consult security champions within their organizational units (see C4 in Table~\ref{keymetrics-champions}). Understanding those reasons will provide deeper insights into how security champion programs influence employee behaviors and areas in which those programs can be improved. 

To measure the effectiveness of security champion programs, the SANS institute proposes comparing the security behaviors of departments (e.g., report rates) that have active security champions to departments that have no security champions (C2). Although this is useful in comparing the effectiveness of the security champion programs across departments as a function of the presence of security champions, it does not provide the ability to compare how security champion density across different departments impacts employee security behavior. Therefore, as illustrated by C3 in Table~\ref{keymetrics-champions}, we propose that measuring the relationship between champion density and security behaviors would provide better insights into how such programs impact the security culture within an organization. For example, based on the correlation coefficient, we can characterize the strength and direction of the linear relationship between the champion density within departments and the  corresponding employee behaviors (e.g., report rates). A positive relationship between champion density and report rates can indicate that the security champion program is effective, and the strength of this relationship would reflect the extent to which such a program improves performance (note though that such a correlational analysis would not provide confirmation of a causal relationship, but would nevertheless support it). Such analysis can be performed on any given security behavior.

\begin{table*}[!h]
  \caption{Employee engagement: metrics and measures to identify behavioral changes of employees}
  \label{keymetrics:engagement}
  \label{tab:observationdetails}
  \centering 
  \resizebox{1\textwidth}{!}{%
 \begin{tabular} { p{.025\textwidth} p{.25\textwidth} p{.55\textwidth}  p{.045\textwidth}   p{.03\textwidth} }
    \toprule
   ID & Name & Description & Existing? & Risk$^*$  \\
    \midrule
    \multicolumn{4}{c}{\textbf{Key metrics}} \\
    \midrule 
  
   E1 & Training completion  &  Percentage of employees who complete mandatory and non-mandatory training   &  Yes & \textcolor{green}{N}  \\
   E2 & Time taken for training &  Average time between training deployment and completion  &  Yes &  \textcolor{green}{N}\\
    E3 & Presentation attendees &  Number of people who attended the security training and awareness sessions & Yes & \textcolor{green}{N}\\
     E4 & Net promoter score (NPS) & The likelihood that an employee would recommend security training and awareness session to others & Yes & \textcolor{green}{N} \\
     E5 & Queries to the security inbox &  The number of queries from employees to the organization's security inbox & Yes &  \textcolor{green}{N} \\
     E6 & Security briefings requests &   Number of requests from employees to the security awareness team to offer security briefings & No & \textcolor{green}{N} \\
  E7 & Hits to the intranet site &  Number of employees reaching out to the intranet for security information & Yes & \textcolor{green}{N}\\
   E8 &    Training started but not completed by the due date   &  Percentage of employees who started the training but failed to complete it before the due date  & No  & \textcolor{red}{P} \\
    E9 &   Number of employees who attended  due to employee referrals &  Number of employees who attended the training and awareness presentations due to  other employees' referrals & No & \textcolor{green}{N}\\
     E10 &  Number of employees who have  provided referrals to other employees  & Number of employees who have referred other employees to  training and awareness presentations &  No &  \textcolor{green}{N}\\
   
     E11 &  Employee social media interactions related to security-related posts  & Qualitative and quantitative analysis of likes, comments and shares  on  social media for security-related posts & No &  \textcolor{green}{N}  \\

    \midrule
    \multicolumn{5}{c}{\textbf{Additional measures and information to better interpret the key metrics}} \\
  
   \midrule 
      E12 &  Reasons for providing low NPS scores & Following up with  employees who provided a low NPS score to understand their reasons for doing so. & Yes & N/A  \\
         E13 &  Reasons for not completing the training & Following up with  employees who did not complete the training to understand the reasons for their behaviors.  & No & N/A \\

    \bottomrule
\end{tabular}}
\smallskip

\parbox[t]{\textwidth}{$^*$If the increase in the metric value is associated with a higher security risk, then we denote this as \textcolor{red}{P}. If the increase in the metric value is associated with a lower security risk, then we denote this as \textcolor{green}{N}. N/A denotes where such direct interpretation is not possible. }
\end{table*}

\subsubsection{Employee engagement}
The more engaged the employees are with the organizational security, the more they will adhere to the organizational security policies and procedures and motivate their colleagues to do the same. Table~\ref{keymetrics:engagement} presents the key metrics and measures that can be used to understand employees’ security related engagements. 

Security training completion (E1) is one key area where employee engagement can be measured. If there is little or no interest in learning and acquiring new skills, unfortunately, creating a good security culture within an organization becomes a challenging task. Non-mandatory training completion rates can vary dramatically  based on the learning culture that exists within an organization. Therefore, it would be useful to report both mandatory and non-mandatory training completion rates separately to obtain more in-depth insights into the organizational security culture. On the other hand, often organizations focus only on the training completion rates; there is less focus on employees who started the training but did not complete it by the due date (E9). Identifying such individuals early will enable organizations to implement measures to remind and motivate those employees to finish their training programs on time. Furthermore, as illustrated in Table~\ref{keymetrics:engagement}, organizations can follow up those employees who did not complete the training to understand the reasons for their actions (E13). Such deeper insights may help organizations to improve the training material and how it is delivered. Furthermore, the longer a user procrastinates completing the assigned training, the less likely they are to complete it. A longer average time between training deployment and completion of a module (E2) could indicate a workforce that does not have the time to prioritize training. 

Organizations usually conduct security awareness presentations for their employees on various topics. To measure the effectiveness of such sessions, organizations often utilize the number of attendees (E3) and the Net Promoter Score (NPS) as key metrics (E4). In the security training and awareness evaluation context, the NPS is an index ranging from -100 to 100 that measures the willingness of employees to recommend a company's training and awareness programs to other employees. This information is often obtained through a survey administrated at the end of a training session. Although NPS is a simple widely used way to measure employees perceived behaviors, there are several limitations with this metric. For example, the literature suggests that intention is not a reliable metric because intentions do not always reflect enacted behavior. Therefore, it is important to complement NPS with other measures to obtain deeper insights into employee involvement in promoting training and awareness programs. Identifying the actual number of employees who attend a training session through referrals (E9) and then following up with them to understand who referred them  (E10) can provide better insights into how well employees actually promote training programs. Furthermore, following up with employees who provide a low value for NPS (E12) can provide detailed insights into where improvements can be made in the security training and education sessions.

\begin{table*}[!h]
  \caption{Protecting organizational and customer information: metrics and measures to identify behavioral changes of employees}
  \label{keymetrics-protecting}
  \label{tab:observationdetails}
  \centering 
  \resizebox{1\textwidth}{!}{%
 \begin{tabular} { p{.025\textwidth} p{.25\textwidth} p{.55\textwidth}  p{.045\textwidth}   p{.03\textwidth} }
    \toprule
   ID & Name & Description & Existing? & Risk$^*$  \\
    \midrule
    \multicolumn{4}{c}{\textbf{Key metrics}} \\
    \midrule 
  
     I1 &  Strength of passwords &  Percentage of employees using strong passwords & Yes & \textcolor{green}{N} \\
      I2 &  Frequency of password change &  Average number of times employees change their passwords &  Yes & \textcolor{green}{N}\\
     I3 &Unsafe storage of passwords & Number of incidents where unsafe storage of passwords was reported& Yes & \textcolor{red}{P}\\
     I4 &  External USB usage & The number of external USB drives used &  Yes &\textcolor{red}{P} \\
     I5 &  Data wiping or deconstruction & Percentage of employees that follow proper the proper protocol for deleting information from retired devices &  Yes & \textcolor{green}{N}\\
      I6 & Accidental data exposure due to auto-complete & Number of accidental data disclosure due to auto-complete in an email.  In other words, the number of employees in the workforce who accidentally email sensitive data to the wrong people &   Yes &\textcolor{red}{P}\\
      I7 & Sensitive data exposure on social media & The number of employees posting sensitive organizational information on social networking sites &   Yes & \textcolor{red}{P}\\
     I8 & Total number of DLP events caused by employee error & Number of escalated Data Loss and Prevention (DLP) events caused by employee error &  Yes &\textcolor{red}{P}\\
      
      I9 & Updated  devices & Percentage of devices that are updated and current & Yes & \textcolor{green}{N}\\
      
 I10 & Installations of a password management tool & The percentage of  employees who have installed a password safe tool  &  No & \textcolor{green}{N}\\
   I11 & Active password management tool (e.g. password safe)  & The percentage of  employees who are actively using  password management tools  &  No &  \textcolor{green}{N}\\
    \midrule
    \multicolumn{5}{c}{\textbf{Additional measures and information to better interpret the key metrics}} \\
  
   \midrule 
          I12 &  Breakdown of the DLP events caused by human errors & Identification of what types of DLP events are caused by human error and their respective frequencies & No & N/A  \\

    \bottomrule
\end{tabular}}
\smallskip

\parbox[t]{\textwidth}{$^*$If the increase in the metric value is associated with a higher security risk,  then we denote this as  \textcolor{red}{P}. If the increase in the metric value is associated with a lower security risk,  then we denote this as \textcolor{green}{N}. N/A denotes where such direct interpretation is not possible. }
\end{table*}

Research suggests that the use of social media in the workplace helps improve employee engagement \cite{wang2016exploring}. In addition to the extension of general social media platforms (e.g., Facebook, Twitter) in the workplace, more specifically social media tools are being implemented in  workplaces, such as Microsoft Yammer and others. We, therefore, suggest that these tools offer opportunities to measure employee behavior in the online environment by analyzing employee interactions with security-related posts on the social media. Analysis of employee engagement on social media over time can provide insights into how behavior changes with time. Both quantitative and qualitative data analysis methods can be utilized in such analysis. For example, sentiment analysis and topic modelling are qualitative data analysis techniques that can be applied to understand the frequency of the discussed topics and emotions associated with those discussions.

\begin{table*}[!h]
  \caption{Physical and device security: metrics and measures to identify behavioral changes of employees}
   \label{tab:physical}
  \centering 
  \resizebox{1\textwidth}{!}{%
 \begin{tabular} { p{.025\textwidth} p{.25\textwidth} p{.55\textwidth}  p{.045\textwidth}   p{.03\textwidth} }
    \toprule
   ID & Name & Description & Existing? & Risk$^*$  \\
    \midrule
    \multicolumn{5}{c}{\textbf{Key metrics}} \\
    \midrule 
    D1 &  Lost and  encrypted   & Percentage of lost and stolen devices that were encrypted & Yes & \textcolor{red}{P}\\ 
     D2 &  Clean desk  & Percentage of  desktops and laptops found unattended without a lock screen  & Yes & \textcolor{red}{P}\\
     D3 &   Secure devices &  Number of devices found unprotected on company property &  Yes & \textcolor{red}{P}\\
  
  

    \bottomrule
\end{tabular}}
\smallskip

\parbox[t]{\textwidth}{$^*$If the increase in the metric value  is associated with a higher security risk then we denote this as \textcolor{red}{P}. If the increase in the metric value is associated with a lower security risk,  then we denote this as \textcolor{green}{N}. N/A denotes where such direct interpretation is not possible.}
\end{table*}

\subsubsection{Protecting organizational and customer information}
Safeguarding organizational and customer information, usually governed by various security policies and privacy laws, is a top priority for an organization. Often security awareness and training programs are designed by organizations to provide awareness about such policies, laws, and best practices to their employees. Table~\ref{keymetrics-protecting} presets metrics to identify related behavioral changes of employees to understand the effectiveness of such programs.

Employees' passwords are gateways to access customer and organizational information; therefore, monitoring how employees use and store passwords is critical. As illustrated in Table~\ref{keymetrics-protecting}, the strength of passwords  (I1) and the frequency with which they are changed (I2) are common ways of measuring employee password-related behaviors. However, during the interviews with our industry partner, it was highlighted that such metrics may not be beneficial as most organizations now employ password protection policies that often make it mandatory for employees to have strong passwords and reset passwords within a given period of time. Indeed, with the increased interest in using password management software, monitoring the percentage of employees who actively use such software  (I10 and I11) is deemed more useful.

Using external USB drives may be considered a risk to an organization's information security (I4). Although it is possible to employ custom software to capture  details about  external hard drive use, it is essential to note that monitoring external hard drive use will depend on the organizational computer security policies. For example, based on such organizational policies, all or certain groups of people may not be allowed to use external devices. In such cases, this metric will have to be used with caution.

Employees distributing sensitive data to unauthorized individuals intentionally or unintentionally is a considerable threat to organizations. Such distribution could expose sensitive organizational and personal data in different ways. This includes but is not limited to employees posting or uploading information to social media and accidentally emailing sensitive data to the wrong recipients. Data Loss Prevention (DLP) or some other perimeter control could be used for monitoring such data exposures. In this regard, most organizations only focus on the total DLP incidents escalated due to human error (I8). We propose also tracking a detailed breakdown of the DLP incidents (I12) caused by human error to obtain further insights into these insider threats. The number of incidents that involve employees uploading or accepting customer information via non-organizational applications and the number of incidents that involve sensitive data being sent to employees' private email addresses are two examples of such metrics. The feasibility of such granular analysis will depend on the organizational tools and processes.

Deleting sensitive information from  devices that a user is retiring demonstrates an understanding of data sensitivity. Therefore, IT teams should capture data about the number of employees who did not follow the proper protocol for deleting information from  retired devices (I5). However, it is important to note that corporate policy around data deletion will affect how much responsibility is put upon users to ensure their devices are wiped.

Part of security training and awareness typically concerns teaching employees about vulnerability and patch management. For the employee, this means keeping personal devices updated. For the business, this means forcing corporate devices to update after a certain period (I9). Corporate policy around patch management and Bring-Your-Own-Devices (BYOD) will affect how much responsibility is put upon users to ensure their devices are updated. Nevertheless, when this responsibility falls upon the user, a useful metric capturing best practice behavior is the percentage of updated and current devices. Information on whether devices are up-to-date or not can be obtained when devices are connected to an internal server or through a third-party vendor.

\subsubsection{Physical and device security}
Employees should be trained in maintaining the physical security of their devices. We identified several metrics to measure how employees are practising these guidelines (see Table~\ref{tab:physical}).

When devices are lost or stolen, it is important for organizations to understand whether or not data on those devices were encrypted (D1). The choice of encrypting their devices demonstrates employees' understanding of the importance of protecting sensitive information. Physical asset audits can identify the number of lost and stolen devices and the percentage that were encrypted. Corporate policy around the encryption of corporate and personal devices will affect how much responsibility is placed upon users to ensure their devices are encrypted.

A clean desk policy (D2) is a corporate directive that specifies how employees should leave their working space when they leave the office. Most clean desk policies require employees to lock their computers when leaving the work place. One way to keep track of who follows the clean desk policy is to inspect the office space to determine if there are unlocked computers.  Furthermore, custom tools  can be designed to collect the security attributes of  employee workstations automatically. Such tools  can reside on employee workstations and report measurements related to various security behaviors of employees (e.g., time of activity without a locked screen) back to its server.  

A secure devices metric (D3) measures the number of devices found unprotected on company property. The data points for this metric often come through  manual tracking via physically walking through of the premises. During a physical walk through of a building and surrounding land (e.g., parking lot or rooftop patio), all devices should be found secured. 
Furthermore,  physical penetration testing, where real-world threat scenarios are simulated  to compromise a business's physical barriers to gain access to infrastructure, buildings, systems, and employees,  could be used to gather data on the overall physical defenses of an organization. 

\begin{figure}[t]
\centering
\includegraphics[width=7.9cm]{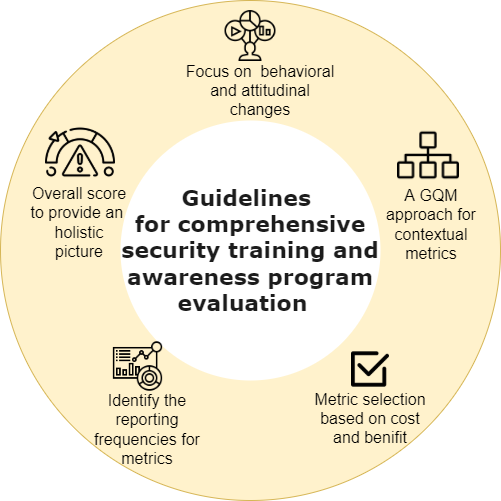}
\caption{Guidelines for security awareness and training program evaluation}
\label{minmap}
\end{figure}

\section{Guidelines for comprehensive  security training  and awareness program evaluation}

The main goal of this work is to provide an overview of practices and metrics for measuring the effectiveness of security training and awareness programs and provide guidelines for a comprehensive evaluation of such programs. In this section, we focus on the latter, where we propose and discuss five guidelines (see Figure~\ref{minmap}) for security training and awareness program evaluation.

\subsection{Focus on  behavioral and attitudinal changes}
Although security training and awareness programs will improve one's security knowledge and security awareness, they cannot guarantee an improvement in behavior. Behavioral scientists have demonstrated that people make decisions based on intuition, emotion, and social pressure in many circumstances, not  on knowledge alone. Therefore, we emphasise that measuring attitudes and behavioral changes is critical in the evaluation of security training and awareness program offered by organizations.

\subsection{A Goal-Question-Metric approach for defining contextual metrics for program evaluation}
For metrics to be meaningful and useful, they must focus on specific goals and be interpreted based on characterisation and understanding of the organizational context, environment, and goals.  Therefore, we propose the use of the Goal Question Metric (GQM) methodology \cite{van2002goal} for systematically defining measurement programs for evaluating the effectiveness of security awareness and training campaigns carried out by organizations. GQM is a well-known methodology for systematically defining effective measurement programs via  establishing goals, formulating questions from the goals, and  designing and performing data measurements based on the questions \cite{van2002goal}.
 The  GQM-based methodology proposed by Philippou et al. \cite{philippou2020contextualising} allows the metrics to be articulated with business goals via preliminary modelling and refinement of such goals, down to a manageable granularity, while also capturing relevant context (business goals, purpose, stakeholders, and system scope) via templates. Furthermore, the GQM model allows feedback from metrics to influence business goals and vice- versa.

\subsection{Metric selection based on cost and benefit}
Defining the metrics based on the organizational goals and the GQM approach alone is not sufficient to conduct a comprehensive evaluation program. It is important to carefully identify the cost involved in data collection and analysis, and the benefit of each of the identified metrics. The key metrics for program evaluation and reporting can be selected by analyzing the trade-off between cost and benefit. The costs and benefits of related metrics for each organization are specific; hence this exercise needs to be carried out by every organization that plans to implement such an evaluation program.

\subsection{Identify the reporting frequencies for metrics}
Since workforce size, relative risk, departmental structure, and numerous other factors relevant to security risks vary between organizations, the frequency with which the selected  metrics  should be collected, analyzed, interpreted, and reported will also vary (e.g., monthly, quarterly, biannually, and annually). For instance, some of the metrics described here are acted upon by behavioral change processes and should be observed for change, whereas others (e.g., personality factors) will fluctuate very little and could therefore be measured less frequently. From our experience with our industry partner, identifying the required reporting frequencies can be an iterative process, and an organization should consider the time and costs associated with calculating metrics when defining the reporting periods. This process may result in sets of metrics that organizations may report with different frequencies. 

\subsection{Overall score to provide an holistic picture}
Through our experience in working with industry over the years, we have learned that often board members or senior executives do not have time to view the details of all the metrics selected in a measurement program. Yet, they are interested in a condensed view that will describe the state of the organization in an easily interpretable manner. Therefore, we suggest calculating a single metric (e.g., human security score) to provide an overview of the effectiveness of cybersecurity training and awareness programs. Having such a standardized way to measure the effectiveness of training and awareness programs allows program comparison at an industry level as well. 

This overall metric can be a summarized form of a few or all of the selected key metrics. Each metric can be converted to a score, grouped under different focus areas and combined to produce a weighted outcome measure of relative risk \cite{SANS}. The importance given to each metric is organization or sector dependent; hence, weights need to be defined at an organizational or sector level. From our experience,  creating metric mockups with real-world data can help organizations to identify how the metric values will look  in practice, have a discussion around the importance of the metric in their security training and awareness programs and decide a weight that should be assigned when calculating the overall score.

\section{Takeaway points}
Increasing security threats are a global concern. 
Often focus is  on deploying technical (e.g., firewalls) countermeasures to prevent security breaches. However, the latest research demonstrates the vulnerability and importance of the human element in the security domain. Employees who are within an organization have legitimate and widespread access and can thus initiate risk purely by making a mistake or unknowingly carrying out an unsafe action. As a result, organizations introduce programs that seek to provide the required security training and awareness for their employees. As raising awareness is an ongoing effort, such campaigns must be evaluated regularly  so that corrective actions can be taken to achieve the best results. 

Through this article, we have presented the state-of-the-art practices and a set of metrics related to security training and awareness program evaluation. We place a special emphasis on understanding behavioral changes of employees in order to identify the effectiveness of and training and awareness program. We have discussed the relevant practices and metrics under five focus areas. Although this article is not intended to provide an exhaustive review of the literature, we believe that we have  provided a broad view of the field by presenting the most prominent practices and metrics for evaluating the effectiveness of security training and awareness programs.   

Although  existing practices and metrics provide much-needed insights into security training and awareness programs, there are several limitations with respect to the identified practices and metrics that demand more work to be undertaken in this space. In an attempt to fill this gap, as a first step, we have provided some suggestions on ways to obtain deeper insights into employees' behavioral changes to better understand the effectiveness of security training and awareness programs. Furthermore, the  guidelines proposed in this article for comprehensive security training and awareness program evaluation can be employed by organizations that wish to initiate or re-structure their security training and awareness programs.

\section{Acknowledgements}
The work has been supported by the Cyber Security Research Centre Limited whose activities are partially funded by the Australian Government’s Cooperative Research Centres Program.
\bibliographystyle{IEEEtran}
\bibliography{Sample-bib}

\end{document}